\documentclass[english,twocolumn,showpacs,preprintnumbers,amsmath,amssymb,prb,aps]{revtex4}
\usepackage[T1]{fontenc}
\usepackage[latin9]{inputenc}
\usepackage{textcomp}
\usepackage{amstext}
\usepackage{graphicx}
\usepackage{amssymb}

\makeatletter

\providecommand{\tabularnewline}{\\}

\@ifundefined{textcolor}{}
{%
 \definecolor{BLACK}{gray}{0}
 \definecolor{WHITE}{gray}{1}
 \definecolor{RED}{rgb}{1,0,0}
 \definecolor{GREEN}{rgb}{0,1,0}
 \definecolor{BLUE}{rgb}{0,0,1}
 \definecolor{CYAN}{cmyk}{1,0,0,0}
 \definecolor{MAGENTA}{cmyk}{0,1,0,0}
 \definecolor{YELLOW}{cmyk}{0,0,1,0}
 }

\usepackage{epsfig}

\makeatother

\usepackage{babel}

\begin{document}

\title{Quantum path integral simulation of isotope effects in the melting
\\  temperature of ice Ih}

\author{R. Ram\'{\i}rez}

\author{C. P. Herrero}

\affiliation{Instituto de Ciencia de Materiales de Madrid (ICMM), 
Consejo Superior
de Investigaciones Cient\'{\i}ficas (CSIC), Campus de Cantoblanco,
28049 Madrid, Spain }
\begin{abstract}
The isotope effect in the melting temperature of ice Ih has been studied
by free energy calculations within the path integral formulation of
statistical mechanics. Free energy differences between isotopes are
related to the dependence of their kinetic energy on the isotope mass.
The water simulations were performed by using the q-TIP4P/F model,
a point charge empirical potential that includes molecular flexibility
and anharmonicity in the OH stretch of the water molecule. The reported
melting temperature at ambient pressure of this model ($T$=251 K)
increases by 6.5$\pm0.5$ K and 8.2$\pm0.5$ K upon isotopic substitution
of hydrogen by deuterium and tritium, respectively. These temperature
shifts are larger than the experimental ones (3.8 K and 4.5 K, respectively).
In the classical limit, the melting temperature is nearly the same
as that for tritiated ice. This unexpected behavior is rationalized
by the coupling between intermolecular interactions and molecular
flexibility. This coupling makes the kinetic energy of the OH stretching
modes larger in the liquid than in the solid phase. However the opposite
behavior is found for intramolecular modes, which display larger kinetic
energy in ice than in liquid water.
\end{abstract}


\maketitle

\section{Introduction\label{sec:intro}}

Quantum mechanical effects associated to the nuclear mass play a significant
role in the properties of liquid water and ice. Experimental evidence
is provided by the isotope dependence of the equilibrium properties
of water. At ambient pressure the melting point at 273.15 K increases
by 3.8 K and 4.5 K after isotopic substitution of hydrogen by deuterium
or tritium, respectively. An even larger isotope effect is found in
the temperature ($T$=277.13 K ) of maximum density (TMD), that increases
at ambient pressure by 7.2 K in heavy water and by 9.4 K in tritiated
water. Such behavior can not be described by classical statistical
mechanics, as in this limit the atomic masses do not affect the phase
diagram of a substance or the equilibrium structure of a liquid. The
importance of quantum effects related to the atomic masses in water
might be expected by the presence of the lightest atom. However, the
larger oxygen mass is also the origin of significant quantum properties
of even practical relevance. For example, the isotopic composition
of the annual layers of ice accumulated in the Antarctica has provided
an indirect measure of the temperature of our planet over the last
400,000 years. The reason is that the vapor pressures of H$_{2}$$^{16}$O,
H$_{2}$$^{18}$O, and D$_{2}$$^{16}$O are different, and then the
isotopic composition of ice results to be a function of the temperature
at which it precipitated. Thus, isotope analysis in ice provides an
historical record for climate change in the past.\citep{petit99}

Computer simulation of water in clusters and condensed phases has
attracted a lot of interest since the pioneering work of Barker and
Watts\citep{barker69} and Rahman and Stillinger\citep{rahman71}
using rigid nonpolarizable models for the water molecule. Since then
a lot of effort has been invested in the development and refinement
of empirical potentials for both water and ice simulations. In fact,
there appears an embarrassing variety of them in the computer simulation
literature. The most employed models assume a rigid geometry of the
water molecule, some include molecular flexibility either with harmonic
or anharmonic OH stretches, and another group deals explicitly with
polarizability effects.\citep{mahoney01} Moreover, in some cases
slight modifications of the potential parameters are proposed for
their use in quantum simulations, to avoid overcounting of quantum
effects if the model parameters were first fitted against experimental
data by classical simulations.\citep{jorgensen05,mcbride09} Besides,
there is an increasing number of water simulations using \textit{ab
initio} density functional theory (DFT).\citep{fernandezserra06,morrone08}
However, the H-bond network, with a strength between weak covalent
and van der Waals interactions, seems difficult to be described with
presently available energy functionals. As a result, some properties
of ice may be poorly reproduced by DFT simulations, e.g., its melting
temperature can be overestimated by more that 130 K.\citep{yoo09}

The melting point of ice at ambient pressure has been determined for
the most common rigid models within the classical limit.\citep{vega05}
It was found that the TIP4P model, with melting point at $T=232$
K, results superior to other models in the sense that correctly predicts
that ice Ih is the stable phase at ambient pressure, while the prediction
of the other rigid models (SPC, SPC/E, TIP3P and TIP5P) was ice II.
An improved parametrization of the rigid model (TIP4P/2005) displays
a melting point at 251 K. \citep{abascal05} Quantum simulations of
phase coexistence are less common than their classical counterparts.
An exception is the work of Habershon \textit{et al}.,\citep{habershon09}
who have developed a flexible water model (q-TIP4P/F) by adding to
the rigid TIP4P/2005 potential the intramolecular flexibility with
the help of OH Morse-type stretches. The model was parametrized on
the basis of quantum path integral (PI) simulations. Its melting point
at 251 K and ambient pressure, was derived by direct coexistence PI
simulations of the water-ice interface. Moreover, in the classical
limit the melting point was found just 8 K above the quantum result.
This temperature shift was considered consistent with the experimental
difference of 4 K between the melting points of H$_{2}$O and D$_{2}$O.\citep{habershon09}
For the point charge flexible q-SPC/Fw model\citep{paesani06} the
classical melting point was however found about 27 K higher than the
quantum result of 195 K.\citep{habershon09} This difference in the
quantum correction of both models might be originated from the description
of the intramolecular OH stretches, i.e., anharmonic (q-TIP4P/F) versus
harmonic (q-SPC/Fw) OH vibrations.

Although the determination of quantum corrections to classical melting
points is interesting because it allows us to quantify the systematic
error of treating water molecules as classical entities, they do not
represent any kind of measurable property. There is no way to perform
measurements of the phase behavior of water in the classical limit.
In this respect, the calculation of the isotope effect in the melting
point of ice offers the advantage of being directly comparable to
experimental data, providing a better test of the capability of the
water model. We are not aware of previous computer simulations of
this isotope effect. However, by the quantum cluster equilibrium theory,
that calculates equilibrium properties by extending standard quantum
statistical thermodynamics of chemical equilibrium to the analogous
equilibrium between molecular clusters, it was estimated that the
melting point of D$_{2}$O is shifted by about 2 K towards higher
temperatures with respect to light water. \citep{ludwig00} Isotope
effects have been studied by PI simulations in many other equilibrium
properties of water. Kuharski and Rossky found that the liquid H$_{2}$O
is less structured than D$_{2}$O. \citep{kuharski85} The explanation
was formulated in terms that the quantum effect associated to the
lower isotope mass results in a less structured H-bond network and
a less tightly bound liquid.\citep{buono95,stern01} Other simulations
of isotope effects focused on the TMD\citep{mahoney01,noya09}, the
diffusion coefficient, \citep{hernandezpena04,paesani07} the heat
capacity, \citep{shiga05,vega10} and the infrared spectra\citep{paesani07}
of water.

In this paper we present a PI simulation of the isotope effect in
the melting temperature of ice Ih at ambient pressure. The q-TIP4P/F
model has been chosen for the simulations because it is an anharmonic
flexible potential whose normal melting point has been already established
by quantum PI simulations.\citep{habershon09} Thus, assuming the
equality of the Gibbs free energy, $G$, of ice Ih and water at the
melting point, the isotope effect will be calculated from the dependence
of $G$ with isotope mass and temperature. Solid-liquid coexistence
will be also studied in the classical limit. The calculation of $G$
will be performed using adiabatic switching (AS) (Ref. \onlinecite {watanabe90})
and reversible scaling (RS) (Ref. \onlinecite {koning99}) approaches,
that are based on algorithms where a Hamiltonian parameter (e.g.,
an atomic mass) or a state variable (e.g., the temperature) changes
along a non-equilibrium simulation run. The capability of both AS
and RS methods to calculate free energies in the context of PI simulations
has been recently analyzed in the study of the phase diagram and isotope
effects of neon.\citep{ramirez08,ramirez08b} 

The structure of this paper is as follows. In Sec. \ref{sec:methodology}
we present the computational conditions employed in the PI simulations
as well as the techniques used to evaluate the free energy as a function
of the isotope mass and temperature. In Sec. \ref{sec:test}, selected
results of quantum simulations are compared to available data of Habershon
\textit{et al.}\citep{habershon09} and also to the classical limit
as a check of the employed computational conditions. In particular,
quantum and classical radial distribution functions (RDFs) of the
liquid phase are presented in Subsec. \ref{subsec:rdf}, while the
quantum and classical TMD of water at ambient pressure is summarized
in Subsec. \ref{subsec:TMD}. The isotope effect in the melting temperature
of ice is the focus of Sec. \ref{sec:isotope}. Results obtained for
D$_{2}$O and T$_{2}$O are compared to available experimental data
in Subsec. \ref{subsec:d2o,t2o}. The classical limit is presented
in Subsec. \ref{subsec:classical-limit}. The calculated isotope effects
are rationalized by a discussion of the mass dependence found for
the kinetic energy (KE) in Subsec. \ref{subsec_kinetic}. Finally,
we summarize our conclusions in Sec. \ref{sec:conclusions}.

\section{methodology\label{sec:methodology}}

\subsection{Computational conditions\label{subsec:pi}}

In the PI formulation of statistical mechanics the partition function
is calculated through a discretization of the integral representing
the density matrix. This discretization defines cyclic paths composed
by a finite number $L$ of steps, that in the numerical simulation
translates into the appearance of $L$ replicas (or beads) of each
quantum particle. Then, the implementation of PI simulations relies
on an isomorphism between the quantum system and a classical one,
derived from the former by replacing each quantum particle (here,
atomic nucleus) by a ring polymer of $L$ classical particles, connected
by harmonic springs with a temperature- and mass-dependent force constant.
Details on this computational method are given elsewhere.\citep{feynman72,gillan88,ceperley95,chakravarty97}
The configuration space of the classical isomorph can be sampled by
a molecular dynamics (MD) algorithm, that has the advantage against
a Monte Carlo method of being more easily parallelizable, an important
fact for efficient use of modern computer architectures. Effective
reversible integrator algorithms to perform PI MD simulations in either
$NVT$ on $NPT$ ensembles ($N$ being the number of particles, $V$
the volume, $P$ the pressure, and $T$ the temperature) have been
described in detail in Refs. \onlinecite{ma96,ma99,tu02,tu98}. Both
isotropic and full cell fluctuations were programmed for the $NPT$
ensemble. All calculations were done using originally developed software
and parallelization was implemented by the MPI library.\citep{pacheco97}

Simulations of water were performed on cubic cells containing 300
molecules, assuming periodic boundary conditions. Ice simulations
included 288 molecules in an orthorhombic simulation cell with parameters
($4a,3\sqrt{3}a,$ 3c), with $(a,c)$ being the standard hexagonal
lattice parameters of ice Ih. A proton disordered ice structure where
each O atom has two chemically bonded and two hydrogen bonded H atoms
with nearly zero dipole moment of the simulation cell was generated
by a Monte Carlo procedure.\citep{buch98} Cell fluctuations in the
extended dynamics of the $NPT$ ensemble were isotropic for the liquid
phase and flexible for the solid simulation cell. The point charge,
flexible q-TIP4P/F model was employed for the simulations.\citep{habershon09}
The Lennard-Jones interaction between oxygen centers was truncated
at $r_{c}$=8.5 \AA, and standard long-range corrections were computed
assuming that the pair correlation function is unity, $g(r)=1$ for
$r>r_{c}$, leading to well-known corrections for the pressure and
internal energies.\citep{johnson93} Long-range electrostatic interactions
and forces were calculated by the Ewald method, and the calculation
was speeded up by allowing the real and reciprocal space sums to be
performed in parallel. The Gaussian smearing parameter in the Ewald
method was set to 0.465 \AA, so that both real and \textbf{\textit{k}}-space
sums require similar computation time. 

To have a nearly constant precision in the PI results at different
temperatures, the number of beads $L$ was set as the integer number
closest to fulfill the relation $LT$=6000 K, i.e., at 300 K the number
of beads was $L=20$. The classical limit is easily achieved within
the PI algorithm by setting $L$=1. The staging transformation for
the bead coordinates was employed for the quantum simulations. Temperature
was controlled by chains of four Nosé-Hoover thermostats coupled to
each of the staging variables, and in the case of the $NPT$ ensemble
an additional chain of four barostats was coupled to the volume.\citep{tu98}
To integrate the equations of motion, a reversible reference system
propagator algorithm (RESPA) was employed.\citep{ma96} For the evolution
of thermostats and harmonic bead interactions a time step $\delta t=\Delta t/4$
was used, where $\Delta t$ is the time step associated to the calculation
of the q-TIP4P/F forces. A value of $\Delta t$=0.3 fs was found to
provide adequate convergence. The virial estimator was employed for
the calculation of the KE,\citep{herman82,parrinello84} and the pressure
estimator was identical to that used in a previous work.\citep{ramirez08c}
Typical runs consisted of $5\times10^{4}$ MD steps (MDS) for equilibration,
followed by runs using between $5\times10^{5}$ and $4\times10^{6}$
MDS for calculation of equilibrium properties. The longest runs were
required for the liquid phase, in particular for the calculation of
the water density as a function of temperature.

\subsection{Relative free energy \label{subsec:freeenergy}}

The thermodynamic integration is a standard technique that allows
us to obtain free energy differences by the calculation of the reversible
work needed to change the original system into a reference state of
known free energy.\citep{kirkwood35,allen,frenkel} The Hamiltonian
is switched along a path that connects both systems and a set of equilibrium
simulations are performed at several points of this path. The AS method
is an alternative procedure where the reversible work is estimated
by slowly changing the system Hamiltonian along a single non-equilibrium
simulation run.\citep{watanabe90} The RS algorithm was formulated
to obtain free energies as a function of a state variable, typically
$T$ or $P$. In this case, a slow (adiabatic) change of the state
variable is performed along the non-equilibrium simulation run.\citep{koning99}
Both AS and RS methods have been recently applied to study the phase
diagram of neon by PI simulations.\citep{ramirez08,ramirez08b} Here
we summarize how these methods fit into our calculation of the isotope
effect in the melting temperature of ice. 

The first step is to obtain the free energy of each phase (solid and
liquid) at some appropriate reference point. The melting temperature
of the q-TIP4P/F model for H$_{2}$O at normal pressure was estimated
by Habershon \textit{et al.} by PI simulations.\citep{habershon09}
Hence this state point ($T_{R}$=251 K, $P_{R}$=1 atm) is a reference
state where the Gibbs free energies of the solid, $G_{s}$, and the
liquid, $G_{l}$, are identical. However, as the melting point was
determined by direct coexistence, the actual value of the free energy
remains unknown. This is not a limitation, as only free energy differences
with respect to the reference state are needed for our purpose of
calculating isotope shifts. In particular, free energy is required
as a function of the hydrogen isotope mass, $m_{F}$, and temperature.
Thus, without loss of generality, we set an arbitrary zero for the
entropy, $S_{0}$, so that we will calculate at ambient pressure relative
free energies, $G_{R}$, defined as\begin{equation}
G_{R}(T,m_{F})=G(T,m_{F})-TS_{0}\;.\label{eq:g_rel}\end{equation}
The pressure dependence of $G$ is omitted here as it is a constant
($P$=1 atm) for all simulations in this work. An alternative to our
choice of setting an arbitrary zero of entropy would be to set an
arbitrary zero for the Gibbs free energy. Both choices would be physically
equivalent in the sense that they would lead to identical results
for the phase coexistence temperature. However, the advantage of setting
a zero of entropy is that then the temperature dependence of $G_{R}$
is given by an expression identical to that valid for $G$ (see Sec.
\ref{subsec:temperature_free_energy }). From the last equation it
is obvious that, at a given temperature, the coexistence condition
of equal free energies of solid and liquid phases, $G_{s}$=$G_{l}$,
implies also that the relative free energies are equal, $G_{R,s}$=$G_{R,l}$.
The arbitrary zero of entropy is formally defined so that the relative
free energy of ice, $G_{R,s}$, (and liquid water $G_{R,l}$) is zero
at the reference state point, $(T_{R},P_{R})$, for water molecules
made of the isotope \textonesuperior{}H with mass $m_{H}$, \begin{equation}
G_{R,s}(T_{R},m_{H})=G_{R,l}(T_{R},m_{H})\equiv0\;.\label{eq:sr}\end{equation}
Our task now is to find the coexistence condition for the other hydrogen
isotopes at ambient pressure, i.e., the temperature, $T_{m}$, that
satisfies equality in the free energies of both solid and liquid phases,\begin{equation}
G_{R,s}(T_{m},m_{F})=G_{R,l}(T_{m},m_{F})\;,\end{equation}
for an isotope mass, $m_{F}$, corresponding either to deuterium ($m_{D}$)
or tritium ($m_{T}$) atoms.

\subsection{Isotope effect on the free energy \label{subsec:isotope_free_energy }}

If temperature is held constant (say at the reference value $T_{R}$),
free energy differences as a function of the isotope mass are the
same, whether calculated with $G$ or $G_{R}$ {[}see Eq. (\ref{eq:g_rel}){]}.
Thus, for simplicity, our derivation is presented here by using $G$
and omitting the indication of its $T$ dependence. Assuming that
the Gibbs free energy, $G(m_{H}),$ of a system made of the isotope
\textonesuperior{}H is known, then the unknown free energy, $G(m_{F})$,
of a system obtained by substituting \textonesuperior{}H by the isotope
of mass $m_{F}$ can be calculated by the expression \begin{equation}
G(m_{F})=G(m_{H})+\intop_{m_{H}}^{m_{F}}\frac{\partial G(m_{I})}{\partial m_{I}}dm_{I}\;.\label{eq:Gm_exact}\end{equation}
We recall the thermodynamic relation between derivatives of the Gibbs
free energy and the Hamiltonian, $\widehat{H}$,\[
\frac{\partial G(m_{I})}{\partial m_{I}}=\left\langle \frac{\partial\widehat{H}}{\partial m_{I}}\right\rangle _{NPT}\;,\]
where the angle brackets represent an ensemble average. The Hamiltonian
associated to the atomic nuclei in the water phase can be expressed
as\begin{equation}
\widehat{H}=\widehat{K}(m_{O})+\widehat{K}(m_{I})+\widehat{V}\end{equation}
where $\widehat{V}$ is the potential energy operator, $\widehat{K}(m_{O})$
represents the sum of the KE operators of all the oxygen nuclei in
the system, while $\widehat{K}(m_{I})$ is the corresponding sum for
the hydrogen isotopes. The mass dependence in the Hamiltonian operator
appears only in the KE operator, thus \begin{equation}
\left\langle \frac{\partial\widehat{H}}{\partial m_{I}}\right\rangle _{NPT}=\left\langle \frac{\partial\widehat{K}(m_{I})}{\partial m_{I}}\right\rangle _{NPT}=-\frac{\left\langle \widehat{K}(m_{I})\right\rangle _{NPT}}{m_{I}}\end{equation}
where the last equality follows from the fact that the mass $m_{I}$
appears in the KE operator as a $m_{I}^{-1}$ factor. $\left\langle \widehat{K}(m_{I})\right\rangle _{NPT}$
is the ensemble average of the sum of the KE of all the hydrogen isotopes
in the system. This quantity is readily obtained in PI simulations
by the virial estimator of the KE, $K(m_{I})$. Inserting the last
result into Eq. (\ref{eq:Gm_exact}), one gets

\begin{equation}
G(m_{F})=G(m_{H})-\intop_{m_{H}}^{m_{F}}\frac{\left\langle K(m_{I})\right\rangle _{NPT}}{m_{I}}dm_{I}\;,\label{eq:Gm}\end{equation}
where $\left\langle K(m_{I})\right\rangle _{NPT}$ is the ensemble
average of the virial KE estimator for the hydrogen isotope. The last
equation shows that the change in the KE as a function of the isotope
mass determines the free energy difference between two isotopes. We
have implemented this free energy evaluation by the AS method.\citep{watanabe90}
The isotope mass, $m_{I}$, is changed from the initial value ($m_{H}$)
to the final one ($m_{F}$) in a single non-equilibrium simulation.
It is convenient to change the integration variable, $m_{I}$, to
the dimensionless parameter, $\lambda_{I}$,\begin{equation}
\lambda_{I}=\frac{m_{H}}{m_{I}}\;,\label{eq:lambda}\end{equation}
so that the integration limits for $\lambda_{I}$ are $\lambda_{H}=1$
and $\lambda_{F}=m_{H}/m_{F}$. For a simulation run consisting of
a total number $J$ of MDS, the actual value of the isotope mass,
$m_{I}$, at the $I$'th simulation step $(I=1,$$\ldots,J)$, is
determined by substituting in Eq.(\ref{eq:lambda}) the following
value of $\lambda_{I}$ \begin{equation}
\lambda_{I}=1+(I-1)\Delta\lambda\;,\end{equation}
where $\Delta\lambda=(\lambda_{F}-\lambda_{H})/(J-1).$ The Gibbs
free energy as a function of the isotope mass $m_{I}$, is discretized
for the simulation steps $I>1$ as\begin{equation}
G(m_{I})=G(m_{I-1})+\frac{1}{2}\left(\frac{K(m_{I})}{\lambda_{I}}+\frac{K(m_{I-1})}{\lambda_{I-1}}\right)\Delta\lambda\;,\label{eq:Gm_AS}\end{equation}
where $K(m_{I})$ is the virial estimator of the KE of the hydrogen
isotope at the simulation step $I$. The change of $\lambda_{I}$
at each simulation step implies to update the corresponding isotope
mass, $m_{I}$, as well as the spring constant for the harmonic coupling
between beads of that isotope. A convenient application of the AS
method is to perform two independent non-equilibrium simulations for
a given free energy determination, where the initial and final integration
limits are interchanged, so that the reversible path between the initial
and final integration points is run in both directions. The free energy
is then obtained as an average of these two independent runs.\citep{ramirez08,ramirez08b}

In the case of the $NVT$ ensemble the mass dependence of the Helmholtz
free energy, $F$, can be derived from the following relation analogous
to Eq. (\ref{eq:Gm})

\begin{equation}
F(m_{F})=F(m_{H})-\intop_{m_{H}}^{m_{F}}\frac{\left\langle K(m_{I})\right\rangle _{NVT}}{m_{I}}dm_{I}\;.\label{eq:Fm}\end{equation}

\subsection{Quantum-classical free energy difference \label{sub:Quantum-classical-free-energy}}

The previous derivation can be applied to calculate the free energy
difference of a classical system with respect to its quantum limit
at the reference state point $(T_{R},P_{R})$. The method implies
the calculation of the reversible work associated to the following
thermodynamic processes in the $NPT$ ensemble: first (process A),
the mass of the water molecules in the quantum system is increased,
so that one formally reaches the limit of infinite molecular mass,
where the classical limit becomes exact; second (process B), the mass
of the water molecules is reduced back to its actual value but now
considering that the system remains in its classical limit. The reversible
work performed along these two processes is the Gibbs free energy
difference between the classical limit and the quantum system. This
work can be calculated by Eq. (\ref{eq:Gm}) if applied to the particular
case where the masses of all the atoms in the system are scaled by
the same factor. It it convenient to define the variable $\lambda$
as

\begin{equation}
\lambda=\frac{M_{0}}{M}\;,\end{equation}
where $M_{0}$ and $M$ are the actual and the scaled molecular masses
of water, respectively. The relative Gibbs free energy difference,
$G_{R,cla}$, of the classical limit with respect to the quantum system
is then

\begin{equation}
G_{R,cla}(M_{0})=\int_{1}^{0}\left(\frac{\left\langle K(M)\right\rangle _{NPT}}{\lambda}-\frac{K_{cla}}{\lambda}\right)d\lambda\;.\label{eq:delta_G_cla}\end{equation}
The first summand in the integral determines the reversible work of
process A, while the second summand corresponds to process B. $\left\langle K(M)\right\rangle _{NPT}$
is the ensemble average of the virial estimator for the total KE (sum
of oxygen and hydrogen atoms) for the quantum system where the molecular
mass of water is $M=M_{0}/\lambda$. $K_{cla}$ is the total KE in
the classical limit that, as a consequence of the equipartition principle,
is independent of the molecular mass $M$ and depends only on temperature
($k_{B}T/2$ per degree of freedom). The integral in Eq. (\ref{eq:delta_G_cla})
can be evaluated by the AS method, i.e., a PI simulation is performed
where the molecular mass is slowly increased as a function of $\lambda.$
The initial mass is $M_{0}$ (set by $\lambda=1$) and the final mass
is a large value defined by an adequate small $\lambda_{F}$ (typically
around 0.01). The value of $G_{R,cla}$ must be then obtained by an
extrapolation of the integral to the value $\lambda_{F}\rightarrow0$,
i.e., 

\begin{equation}
G_{R,cla}=\lim_{\lambda_{F}\rightarrow0}I(\lambda_{F})\;,\label{eq:I_lim}\end{equation}
where $I(\lambda_{F})$ is the definite integral in Eq. (\ref{eq:delta_G_cla})
when the upper integration limit is $\lambda_{F}$ instead of zero.
In Appendix \ref{appendix: harmonic} it is shown how the harmonic
result for Eq. (\ref{eq:delta_G_cla}) let us expect that the extrapolation
can be accurately done by a simple polynomial fit in the variable
$\lambda_{F}$. An alternative calculation of the free energy, $G_{R,cla}$,
is the method of Morales and Singer,\citep{morales91} based on a
thermodynamic integration that depends only on the potential energy.
However, this method is not so convenient for the calculation of free
energy differences between isotopes, as it would require to use the
classical limit as an intermediate step to set up a reversible path
connecting both isotopes.

\subsection{Temperature dependence of the free energy\label{subsec:temperature_free_energy }}

The determination of the solid-liquid coexistence temperature requires
to calculate the free energy of both phases over an overlapping range
of temperatures. To this aim we consider the following thermodynamic
relation that relates the enthalpy, $H(T)$, with the Gibbs free energy

\begin{equation}
H(T)=\frac{\partial\left[\beta G(T)\right]}{\partial\beta}\;,\label{eq:enthalpy}\end{equation}
where $\beta=1/k_{B}T$ is the inverse temperature. Here, we have
omitted the explicit indication of the pressure and mass dependence
of $G$ and $H$, that are assumed to be constant. Integrating this
equation between the inverse reference temperature, $\beta_{R}$,
and a final inverse temperature, $\beta_{F}$, one gets

\begin{equation}
\beta_{F}G(T_{F})=\beta_{R}G(T_{R})+\int_{\beta_{R}}^{\beta_{F}}d\beta H(T)\;.\label{eq:g/t}\end{equation}
The RS method is implemented here by a non-equilibrium simulation
where the inverse temperature is changed uniformly along a simulation
run composed of $J$ MDS as

\begin{equation}
\beta_{i}=\beta_{R}+(i-1)\Delta\beta\;,\end{equation}
where $\beta_{i}$ is the inverse temperature of the $i'$th simulation
step ($i=1,\ldots,J$) and the increment is $\Delta\beta=(\beta_{F}-\beta_{R})/(J-1)$.
The free energy at temperature $T_{i}$ is discretized from Eq. (\ref{eq:g/t})
as \begin{equation}
\beta_{i}G(T_{i})=\beta_{i-1}G(T_{i-1})+\Delta\beta\left[\frac{H(T_{i})+H(T_{i-1})}{2}\right]\;.\label{eq:g_i/t_i}\end{equation}
$H(T_{i})$ is the system enthalpy at the $i$'th simulation step.
This expression corresponds to the $NPT$ ensemble and is also valid
if $G$ is substituted by the relative energy $G_{R}$, because $\beta G$
and $\beta G_{R}$ differ only in a temperature-independent constant,
$S_{0}$ {[}see Eq. (\ref{eq:g_rel}){]}. The change of the inverse
temperature $\beta_{i}$ at each simulation step implies to update
all variables that depend on temperature in our PI MD algorithm. These
variables are the spring constants between beads, as well as the thermostat,
barostat, and volume masses defined for the extended dynamics in the
$NPT$ ensemble. \citep{ma99,tu02,tu98}

A relation equivalent to Eq. (\ref{eq:g_i/t_i}) can be derived in
the $NVT$ ensemble if $G$ is substituted by the Helmholtz free energy,
$F$, and the enthalpy, $H$, by the internal energy $U$.

\section{Test simulations\label{sec:test}}

\subsection{Radial distribution function\label{subsec:rdf}}

\begin{figure}[!t]
\vspace{-1.0cm}
\hspace*{-0.1cm}
\includegraphics[width= 9cm]{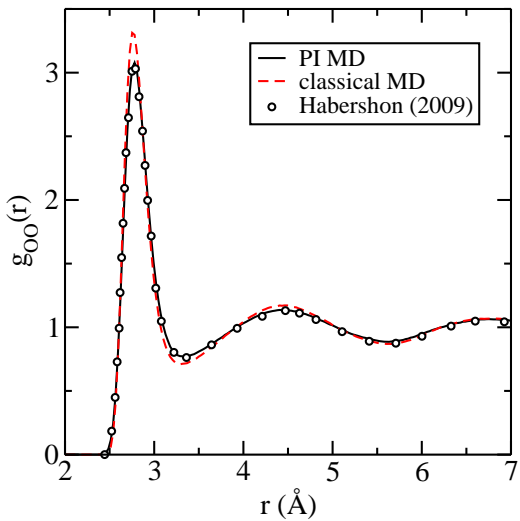}
\vspace{-1.3cm}
\caption{OO RDFs derived from quantum and classical simulations of
water at
298 K and density 0.997 g cm$^{-3}$. For comparison the PI MD results
of Ref. \onlinecite{habershon09} are shown as open circles.}
\label{fig:1}
\end{figure}

\begin{figure}
\vspace{-0.6cm}
\hspace*{-0.1cm}
\includegraphics[width= 9cm]{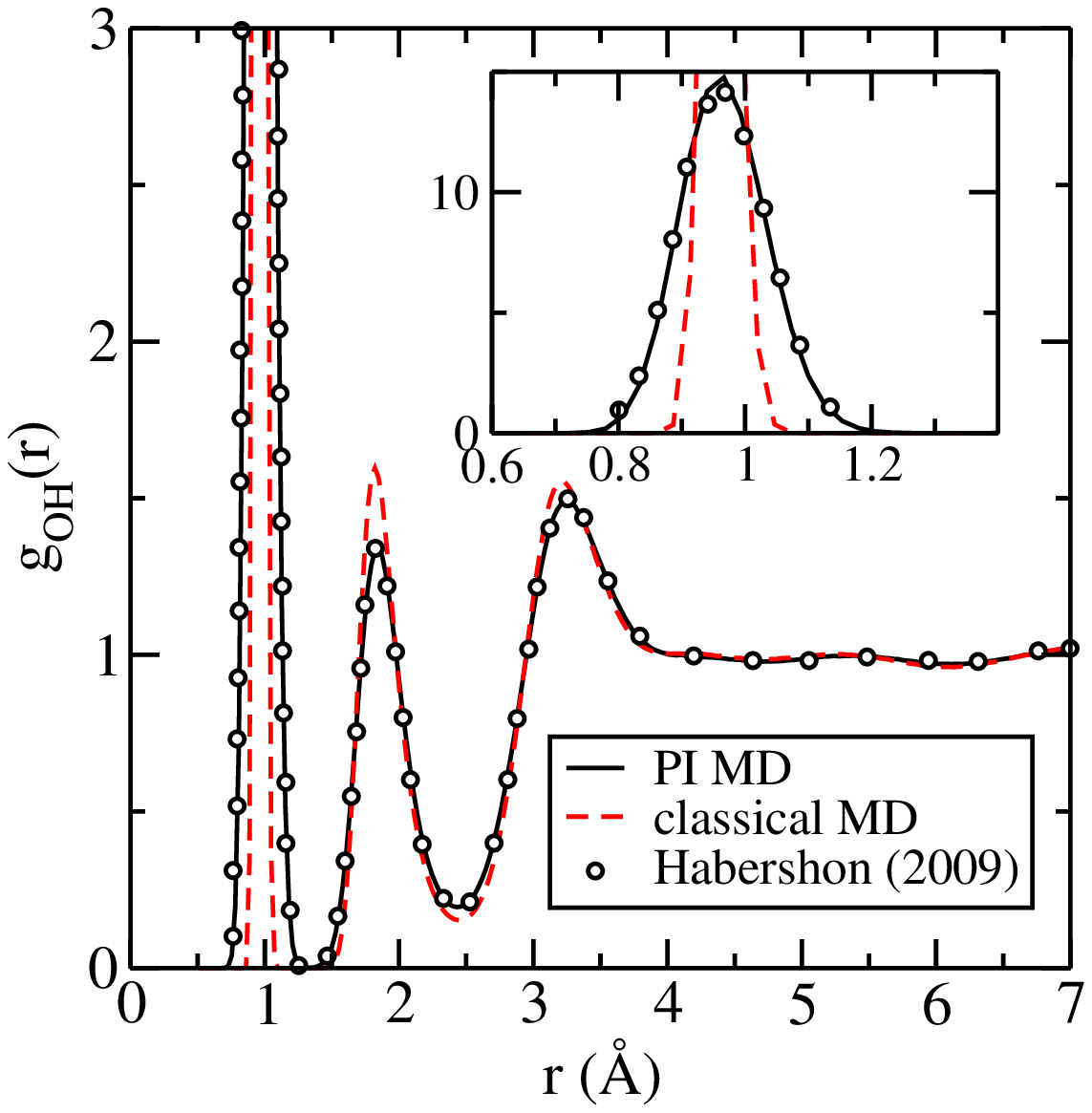}
\vspace{-1.3cm}
\caption{OH RDFs derived from quantum and classical simulations of
water at
298 K and density 0.997 g cm$^{-3}$. For comparison the PI MD results
of Ref. \onlinecite{habershon09} are shown as open circles.}
\label{fig:2}
\end{figure}

\begin{figure}
\vspace{-1.0cm}
\hspace*{-0.1cm}
\includegraphics[width= 9cm]{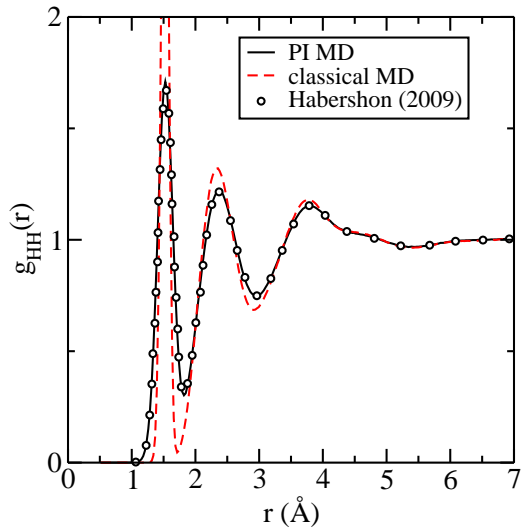}
\vspace{-1.3cm}
\caption{HH RDFs derived from quantum and classical simulations of
water at
298 K and density 0.997 g cm$^{-3}$. For comparison the PI MD results
of Ref. \onlinecite{habershon09} are shown as open circles.}
\label{fig:3}
\end{figure}

The RDFs for OO, OH, and HH pairs have been calculated for water in
the $NVT$ ensemble at 298 K and density 0.997 g cm$^{-3}$. The results
of quantum and classical simulations are presented in Figs. \ref{fig:1}
to \ref{fig:3}. The curves were derived from runs composed of 10$^{6}$
MDS. Our quantum results have been compared to those published in
Ref. \onlinecite{habershon09}. We observe in the three RDFs an excellent
agreement between both sets of independent calculations. The technical
setup of the simulations differs in many details, as the number of
employed beads, the use of staging versus normal mode representations
of the bead coordinates, the thermostats setup, the cutoff distance
for short-range interactions, the approximation used to perform the
Ewald summation in reciprocal space,\citep{habershon09} and the schemes
used for the RESPA molecular dynamics. Therefore, the agreement found
by the RDFs curves provides a check for the accuracy of the employed
computational conditions.

\begin{table}
\caption{Molecular properties (bond distance, bond angle and dipole
moment)
as well as kinetic ($K$) and potential energy ($U_{pot}$) of liquid
water at 298 K and density $\rho=0.997$$\;$g cm$^{-3}$. $r_{O\cdots H}$
is the distance of the $g_{OH}$ RDF maximum associated to the H-bond.
The KE is partitioned into H-isotope and O-atom contributions ($K_{I}$
and $K_{O}$, respectively). $\rho_{TMD}$ is the maximum density
of water at TMD, as derived from $NPT$ simulations at $P$ = 1 atm.
Both classical and quantum results are given. The quantum results
correspond to normal (H$_{2}$O), heavy (D$_{2}$O), and tritiated
(T$_{2}$O) water.}
\centering{}\label{tab:1}\begin{tabular}{lccccc}
\hline
 & classical & T$_{2}$O & D$_{2}$O  & H$_{2}$O  & H$_{2}$O %
\footnote{Ref. {[}\onlinecite{habershon09}{]}%
}\tabularnewline
\hline
$r_{O\cdots H}\;$(\AA) & \multicolumn{1}{c}{1.82} & 1.83 & 1.83 & 1.84
& 1.84\tabularnewline
$\left\langle r_{OH}\right\rangle \;$(\AA) & 0.963 & 0.972 & 0.974 &
0.977 & -\tabularnewline
$\left\langle \theta_{HOH}\right\rangle \;$(deg) & 104.8 & 104.8 &
104.7 & 104.7 & -\tabularnewline
$\left\langle \mu\right\rangle \;$(D) & 2.312 & 2.333 & 2.338 & 2.346 &
-\tabularnewline
$\left\langle K\right\rangle \;$(kcal mol$^{-1}$) & 2.68 & 5.78 & 6.51
& 8.13 & -\tabularnewline
$\left\langle K_{I}\right\rangle \;$(kcal mol$^{-1}$) & 1.79 & 4.35 &
5.14 & 6.86 & -\tabularnewline
$\left\langle K_{O}\right\rangle \;$(kcal mol$^{-1}$) & 0.89 & 1.43 &
1.37 & 1.27 & -\tabularnewline
$\left\langle U_{pot}\right\rangle \;$(kcal mol$^{-1}$) & -10.31 &
-7.13 & -6.39 & -4.57 & -\tabularnewline
\hline
TMD (K) & 282(2) & - & - & 280(2) & 279(2)\tabularnewline
$\rho_{TMD}$ (g cm$^{-3}$) & 1.004(2) & - & - & 1.002(2) &
1.001(2)\tabularnewline
\hline
\end{tabular}
\end{table}

The comparison of the RDFs obtained by PI simulations with the q-TIP4P/F
model to experimental curves has been presented elsewhere\citep{habershon09}
and will not be repeated here. However, we focus here on differences
found between quantum and classical limits of the RDFs of this model.
The OO RDF curves in Fig. \ref{fig:1} show that classical water is
more structured than in the quantum case. In the classical limit,
the height of the first OO peak is larger and the position of the
maximum is displaced by about 0.01 \AA $\;$towards shorter OO distances.
The second peak in the OH RDF curves in Fig. \ref{fig:2} corresponds
to the H-bond distances. Controversial quantum and classical results
have been published for this peak. A DFT study shows that in a quantum
simulation this peak appears at shorter distances, which was interpreted
as a hardening of the water structure with respect to classical simulations.\citep{chen03}
However, most simulations predict that quantum water is less structured
than the classical counterpart.\citep{morrone08} Our quantum results
for the q-TIP4P/F model show that this peak appears at a distance
0.02 \AA\ larger than in the classical limit (see Table \ref{tab:1}),
in agreement to the expectation that quantum corrections destabilize
the H-bond network. 

A comparison of other equilibrium results of quantum and classical
water simulations is summarized in Table \ref{tab:1}. The quantum
results have been derived for both normal, heavy and tritiated water.
We see that quantum corrections associated to the atomic mass increase
the intramolecular OH bond length by more than 0.01 \AA, and the
average molecular dipole moment, $\left\langle \mu\right\rangle $,
by 1.5\%. Note that this increase in $\left\langle \mu\right\rangle $
is expected to act against the destabilization of the H-bond network,
as the electrostatic interaction between neighboring water molecules
becomes stronger. In Tab. \ref{tab:1} we also compare the average
kinetic and potential energies of water. We note that the KE of normal
water is 5.45 kcal mol$^{-1}$ larger in the quantum case. This increase
is even larger for the potential energy (5.74 kcal mol$^{-1}$) as
a consequence of the anharmonicity of the model potential. A harmonic
approximation predicts that both kinetic and potential energy increments
must be identical. The partition of the KE between H- and O-atom contributions
shows that the three-fold rise of the KE of normal water in the quantum
case is mainly due to the H-atoms. For comparison we note that $^{20}$Ne
is the Lennard-Jones-type system whose atomic mass is closest to that
of a water molecule. For solid $^{20}$Ne at 24 K (a temperature close
to the melting point at atmospheric pressure) the rise in KE amounts
to a factor of about 1.4 with respect to the classical limit.\citep{herrero01}

\subsection{Temperature of maximum density\label{subsec:TMD}}

\begin{figure}
\vspace{-1.9cm}
\hspace*{-0.1cm}
\includegraphics[width= 9cm]{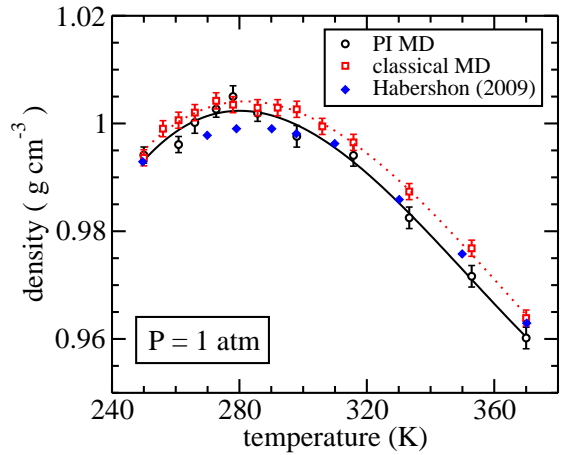}
\vspace{-1.3cm}
\caption{Density of water at 1 atm pressure obtained from classical and
PI MD simulations. Lines are cubic polynomial fits to the simulation
data.}
\label{fig:4}
\end{figure}

We have performed quantum and classical $NPT$ simulations to calculate
the temperature dependence of the water density, $\rho,$ at ambient
pressure. The results are summarized in Fig. \ref{fig:4}. Our quantum
results, shown by open circles, are again in reasonable agreement
to the PI MD data of Ref. \onlinecite{habershon09} for the same model
potential. The densities calculated by both sets of independent simulations
differ by not more than 0.5\%, which is of the order of the estimated
statistical error. A systematic deviation is found at temperatures
corresponding to the maximum density, indicating that the convergence
over density fluctuations is particularly difficult in this region.
The TMD, as obtained by a cubic polynomial fit to the simulation points,
is found at 280 K, in good agreement to previously published results
for this potential (see last two rows in Table \ref{tab:1}).\citep{habershon09}
A comparison of simulation results to experiment has been presented
elsewhere,\citep{habershon09} showing that the employed model provides
realistic results. We recall that a proper description of the density
of water is an important requirement for a water model to account
for hydration effects.\citep{paschek04} The classical TMD is found
at slightly higher temperature than in the quantum case, the shift
amounts to 2 K, which is of the order of the statistical error. Classical
and quantum temperature-density curves are so close that we have not
tried to determine the hydrogen isotope effect in the TMD. However,
we have checked that the density in simulations with tritiated water
is, within the statistical error, very close to the normal water results.
Therefore the employed q-TIP4P/F potential seems to be unable to reproduce
the experimental isotope shift of the TMD in water.

The small difference between the classical and quantum TMD is consistent
with the previous work of Habershon \emph{et al.\citep{habershon09}
}who found similar trends by comparing classical and quantum diffusion
coefficients. A different result was derived for a rigid water model,
TIP4PQ/2005. For this model, the classical TMD was found about 27
K higher than the quantum temperature, and the isotope effect for
tritiated water amounts to 16 K, overestimating the experimental result
of 9.4 K.\citep{noya09} The main difference between the rigid TIP4PQ/2005
and the flexible q-TIP4P/F potential is the presence of an anharmonic
intramolecular potential in the latter. The mechanism that has been
previously used to explain the striking differences between quantum
and classical diffusion coefficient of water when using either anharmonic
or rigid intramolecular models,\citep{habershon09} can be also used
here to explain the differences found in the quantum and classical
TMDs when using either anharmonic or rigid models. This mechanism
implies the coupling between anharmonic intramolecular stretches (OH)
and intermolecular H-bonds (O$\cdots$H) in water. Anharmonic zero
point effects weaken both O$\cdots$H and OH bonds (see the comparison
of distances $\left\langle r_{OH}\right\rangle $ and $r_{O\cdots H}$
in Table \ref{tab:1}, quantum results are larger than classical ones).
However, weakening of the OH bond increases the molecular dipole moment,
that is accompanied by a concomitant strengthening of the intermolecular
H-bond. Therefore, when quantum effects associated to the nuclear
masses are included in an anharmonic flexible water potential, the
H-bond is affected by competing factors acting in opposite directions,
that might lead even to a mutual cancellation: $a)$ weakening by
zero point effects; $b)$ strengthening by the increase of the molecular
dipole moment. Obviously this competing mechanism is absent if the
water model is rigid. Which of the two competing effects is dominant
will depend on the details of the potential model and on the physical
property under consideration. For the TMD of the q-TIP4P/F model,
it seems that both competing effects cancel each other, leading to
nearly identical quantum and classical curves of the liquid density
as a function of temperature and then to a vanishingly small isotope
effect in the TMD of water.

\section{isotope effects in the melting temperature\label{sec:isotope}}

\subsection{D$_{2}$O and T$_{2}$O\label{subsec:d2o,t2o}}

\begin{table}
\caption{Relative free energies, $G_{R}$, of tritiated water at the
reference
point ($T_{R,}P_{R}$) as derived by independent AS simulations of
different lengths. For a given simulation length two results are
presented,
corresponding to simulations where the initial and final integration
limits ($m_{H,}m_{T}$) are interchanged. The last column shows the
average of both independent runs.}
\label{tab:2}
\begin{tabular}{c|ccc}
\hline
MDS &  & $G_{R}$ (kcal mol$^{-1}$) & \tabularnewline
\cline{1-1}
 & $m_{H}\rightarrow m_{T}$ & $m_{T}\rightarrow m_{H}$ &
average\tabularnewline
\cline{2-4}
10$^{5}$ & -5.947 & -5.950 & -5.949\tabularnewline
2$\times10^{5}$ & -5.951 & -5.949 & -5.950\tabularnewline
4$\times10^{5}$ & -5.949 & -5.951 & -5.950\tabularnewline
1.5$\times10^{6}$ & -5.950 & -5.952 & -5.951\tabularnewline
\hline
\end{tabular}
\end{table}

\begin{figure}
\vspace{-1.8cm}
\hspace*{-0.1cm}
\includegraphics[width= 9cm]{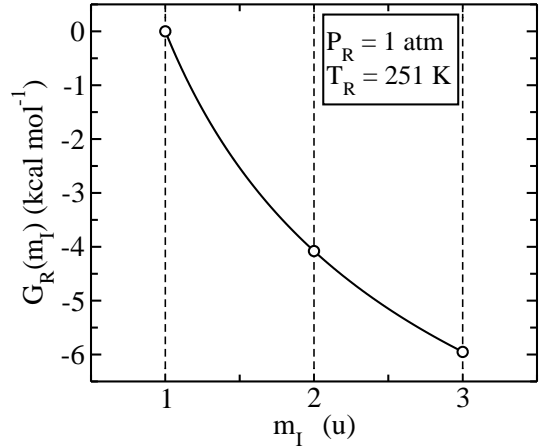}
\vspace{-1.3cm}
\caption{Relative free energy of liquid water as a function of the
hydrogen
isotope mass. The open circles corresponds to the masses of H, D,
and T, respectively. The result was obtained by non-equilibrium
simulations
with the AS method at the reference state point ($T_{R},P_{R}$).}
\label{fig:5}
\end{figure}

To study the isotope effect in the melting temperature of ice we need
first to calculate the relative free energies, $G_{R}$, of deuterated
and tritiated phases at the reference state point ($T_{R},P_{R}$).
To this aim we have performed AS simulations in the $NPT$ ensemble
according to Eq. (\ref{eq:Gm_AS}). For the liquid phase, the dependence
of $G_{R}$ with the isotope mass is shown in Fig. \ref{fig:5}. The
convergence of $G_{R}$ has been checked by comparing independent
non-equilibrium simulations of different lengths for both solid and
liquid phases. This test is presented in Table \ref{tab:2} for tritiated
water. We find that the statistical error in the free energy is even
lower for heavy water and for the solid phases. Even with a modest
number of simulation steps ($J=10^{5}$) the AS method shows reasonable
convergence. Our final results for $G_{R}$ are summarized in Table
\ref{tab:3}, corresponding to non-equilibrium simulations with 
1.5$\times10^{6}$
MDS for the liquid and 4 $\times10^{5}$ MDS for the solid phases.
For the heavier isotopes (D$_{2}$O and T$_{2}$O) the relative free
energy of the solid at the reference state point is lower than in
the liquid phase. This difference determines an isotope shift in the
melting temperature, as temperature must be increased to restore the
coexistence condition of equal free energy for both phases. The free
energy difference between solid and liquid is larger for tritium than
for deuterium, therefore the isotope shift in the melting temperature
is expected to be larger for tritiated water than for deuterated water.

\begin{table}
\caption{Relative free energy, $G_{R}$, at the reference point
$(T_{R}=251$
K, $P_{R}$= 1 atm) of solid and liquid phases of normal, heavy, and
tritiated water. $G_{R}$ is given in kcal mol$^{-1}$, and its estimated
error is $\pm0.001$ kcal mol$^{-1}$. The last column summarizes
the results obtained in the classical limit ($G_{R,cla}$).}
\label{tab:3}
\begin{tabular}{ccccc}
\hline
 & H$_{2}$O & D$_{2}$O & T$_{2}$O & classical limit\tabularnewline
\hline
solid (s) & 0 & -4.108 & -5.987 & -8.912\tabularnewline
liquid (l) & 0 & -4.079 & -5.951 & -8.875\tabularnewline
s $-$ l  & 0 & -0.028 & -0.036 & -0.036\tabularnewline
\hline
\end{tabular}
\end{table}

\begin{figure}
\vspace{-1.8cm}
\hspace*{-0.1cm}
\includegraphics[width= 9cm]{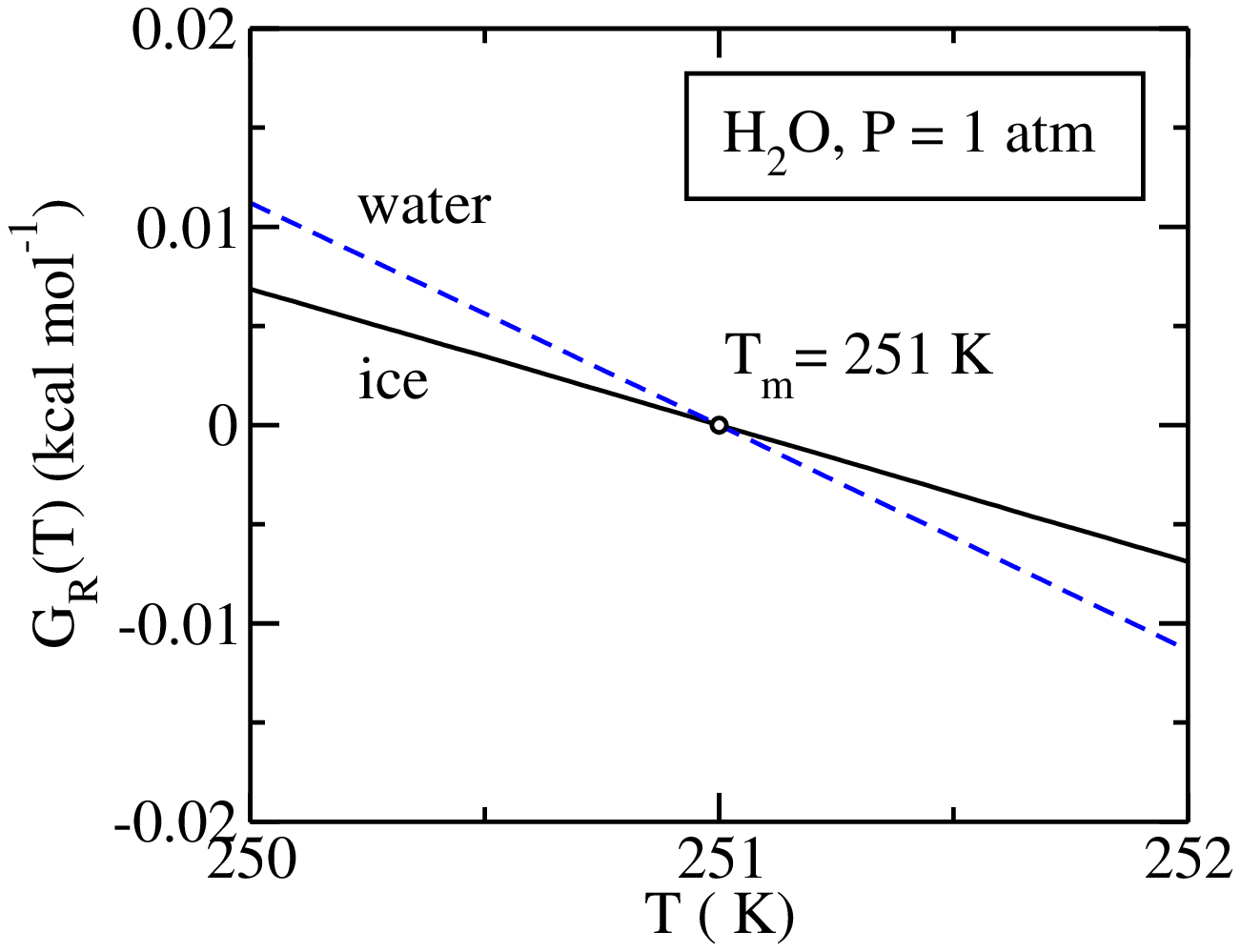}
\vspace{-1.3cm}
\caption{Relative free energy of normal water and ice at pressure of 1
atm
as determined by our RS simulations. The melting point is $T_{m}$.}
\label{fig:6}
\end{figure}

\begin{figure}
\vspace{-0.6cm}
\hspace*{-0.1cm}
\includegraphics[width= 9cm]{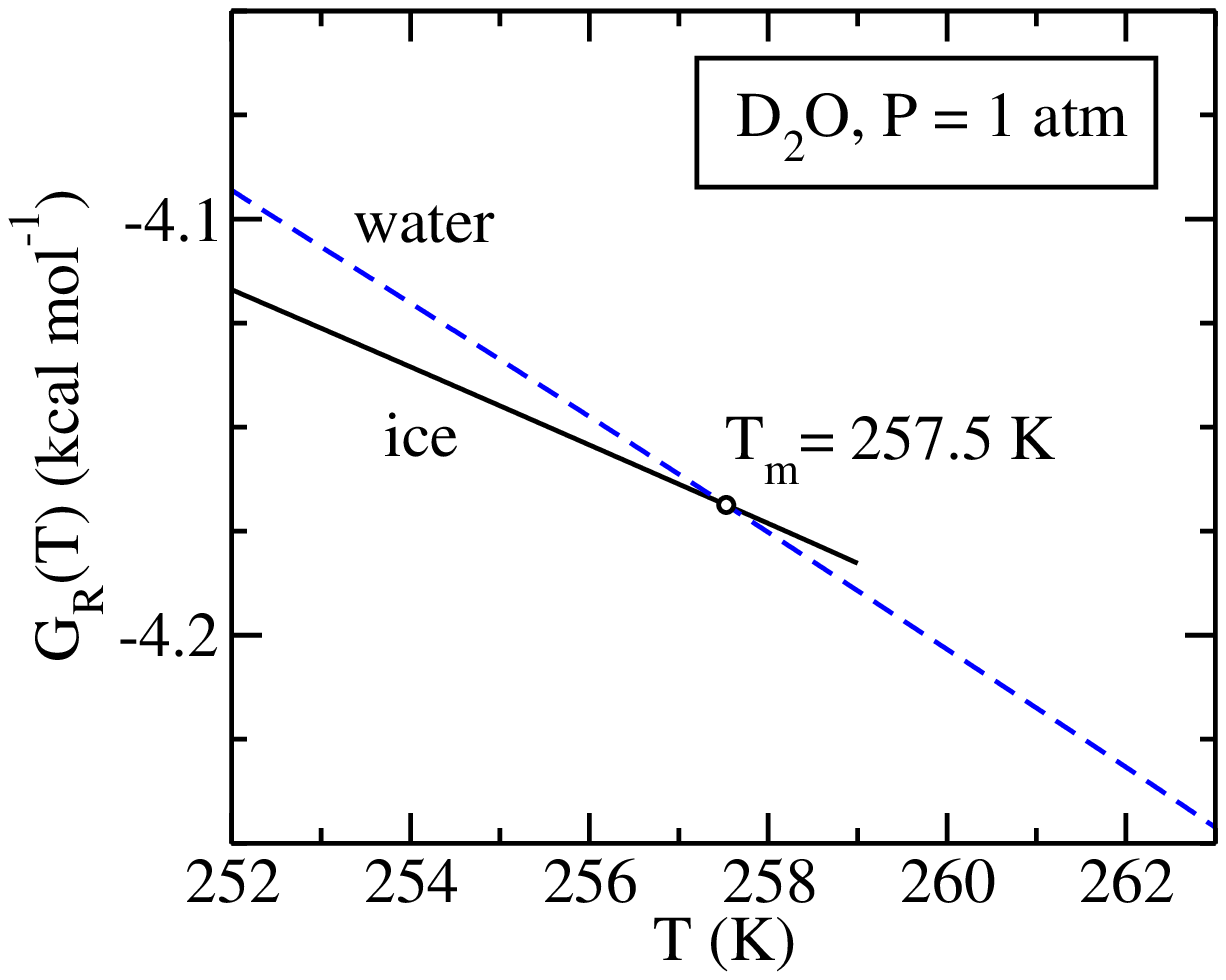}
\vspace{-1.3cm}
\caption{Relative free energy of deuterated water and ice at pressure
of 1
atm as determined by our RS simulations. The melting point is $T_{m}$.}
\label{fig:7}
\end{figure}

\begin{figure}
\vspace{-1.9cm}
\hspace*{-0.1cm}
\includegraphics[width= 9cm]{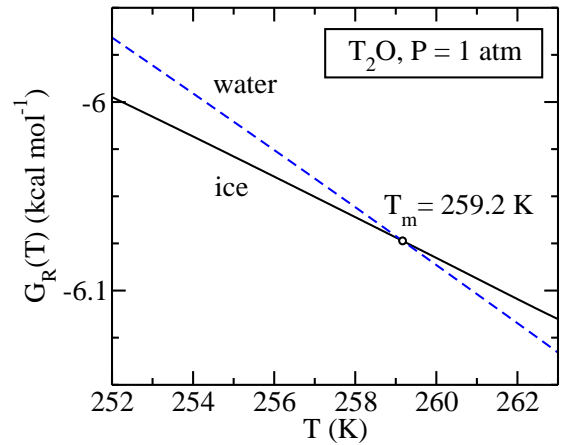}
\vspace{-1.3cm}
\caption{Relative free energy of tritiated water and ice at pressure of
1 atm
as determined by our RS simulations. The melting point is $T_{m}$.}
\label{fig:8}
\end{figure}

The temperature dependence of $G_{R}$ has been calculated by RS simulations
in the $NPT$ ensemble {[}see Eq.(\ref{eq:g_i/t_i}){]}. Several independent
simulations of different lengths were performed to fix the conditions
of adequate convergence for the non-equilibrium simulations. Again,
each free energy estimation was based on two independent simulations
where the integration limits of the temperature in the reversible
path were interchanged, and the final results were obtained as an
average of both RS simulations. The employed computational conditions
(integration limits for the temperature and number of MDS) used in
our RS simulations are summarized in Table \ref{tab:4}. The liquid
RS simulations require typically simulation lengths one order of magnitude
larger than the solid phase. The $G_{R}(T)$ results obtained for
normal, deuterated and tritiated phases are presented in Figs. \ref{fig:6}-\ref{fig:8}.
The condition of equal free energy of the solid and liquid phases
determines the melting point at ambient pressure. The melting temperature
is shifted towards higher values as the hydrogen isotope mass increases.
The estimated isotope shift in the melting temperature is 6.5$\pm$0.5
K for heavy water and 8.2$\pm$0.5 K for tritiated water for the q-TIP4P/F
model. Experimental isotope shifts amount to 3.8 K and 4.5 K, respectively.
Thus, the q-TIP4P/F model overestimates the isotope shift in the melting
temperature of ice. This result contrasts with the nearly absence
of isotope shifts in the TMD of water, as predicted by the same potential. 

\begin{table}
\caption{Computational conditions used in the non-equilibrium RS
simulations
to determine the temperature dependence of the relative Gibbs free
energy of the studied isotope compositions of water and ice. }
\label{tab:4}
\begin{tabular}{cccc}
\hline
isotope & phase & $T$ range (K)  & MDS\tabularnewline
\hline
H$_{2}$O & liquid & 250-252 & 2$\times10^{5}$\tabularnewline
H$_{2}$O & solid & 250-252 & 5$\times10^{4}$\tabularnewline
D$_{2}$O & liquid & 251-263 & 2$\times10^{6}$\tabularnewline
D$_{2}$O & solid & 251-259 & 2$\times10^{5}$\tabularnewline
T$_{2}$O & liquid & 251-263 & 2$\times10^{6}$\tabularnewline
T$_{2}$O & solid & 251-263 & 2$\times10^{5}$\tabularnewline
classical & liquid & 251-261 & 3$\times10^{6}$\tabularnewline
classical & solid & 251-273 & 8$\times10^{5}$\tabularnewline
\hline
\end{tabular}
\end{table}

Melting in atomic solids can be thought of as being controlled by
the local vibrations of the atoms. The Lindemann criterion presents
a threshold for the maximum amplitude of atomic vibrations that can
be sustained by the crystal. At the melting point of ice we find that
this amplitude is similar for both O- and H-atoms and amounts to about
6\% of the H-bond distance.

\subsection{Classical limit\label{subsec:classical-limit} }

\begin{figure}
\vspace{-1.5cm}
\hspace*{-0.1cm}
\includegraphics[width= 9cm]{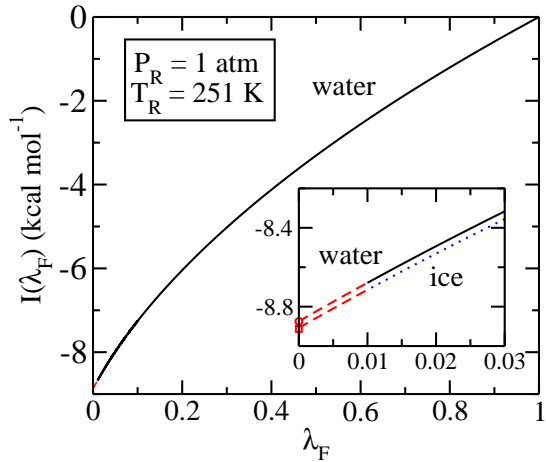}
\vspace{-1.3cm}
\caption{The function $I(\lambda_{F})$ obtained for liquid water by
non-equilibrium
AS $NPT$ simulations at the reference state point ($T_{R},P_{R}$).
The extrapolation $\lambda_{F}\rightarrow0$ is shown for the solid
and liquid phases in the inset of the figure.}
\label{fig:9}
\end{figure}

\begin{figure}
\vspace{-0.3cm}
\hspace*{-0.1cm}
\includegraphics[width= 9cm]{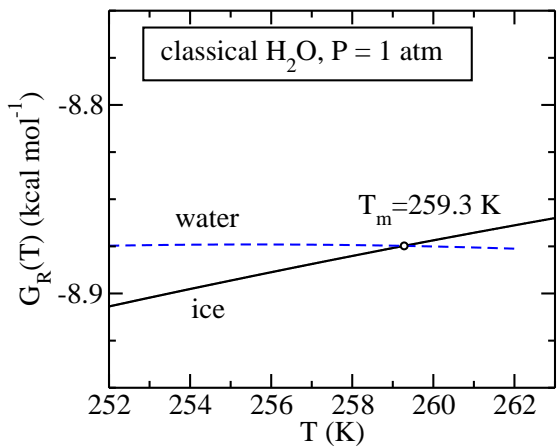}
\vspace{-1.3cm}
\caption{Relative free energy of water and ice at pressure of 1 atm as
determined by our RS simulations in the classical limit. The melting
point is
$T_{m}$.}
\label{fig:10}
\end{figure}

The determination of the classical melting point implies to calculate
the relative free energy, $G_{R,cla}$, of the classical limit of
water and ice with respect to the quantum case at the reference state
point. To this aim we have performed AS simulations where the molecular
mass of water is slowly changed from its normal value, $M_{0},$ to
an arbitrary large value, $M_{F}$, determined by the parameter $\lambda_{F}=M_{0}/M_{F}$.
The function $I(\lambda_{F})$ introduced in Sec. \ref{sub:Quantum-classical-free-energy}
{[}see Eq. (\ref{eq:I_lim}){]} is then obtained by a discretization
of the integral given in Eq. (\ref{eq:delta_G_cla}) as a running
average along the non-equilibrium simulation. The AS results in the
solid phase have been derived with 4$\times10^{5}$ MDS while the
liquid phase required longer runs of 1.6$\times10^{6}$ MDS. As usual
in our non-equilibrium simulations, two independent simulation runs
were performed by interchanging the integration limits, and the final
free energy is the average of these two runs. The function $I(\lambda_{F})$
for the liquid phase is shown in Fig. \ref{fig:9}. The inset in the
figure shows the result of the extrapolation $\lambda_{F}\rightarrow0$
for both solid and liquid phases. The extrapolated values are summarized
in the last column of Table \ref{tab:3}. These figures were obtained
by a polynomial fit of $I(\lambda_{F})$, after checking that the
extrapolation is reasonably stable against both the polynomial degree
and the $\lambda_{F}$ interval employed in the numerical fit. The
results given in Table \ref{tab:3} were derived by a 10'th degree
polynomial fit in the $\lambda_{F}$ range {[}0.01,0.2{]}. Somewhat
surprising is that the difference between the relative free energy
of the solid and liquid phases in the classical limit is, within the
statistical error, identical to the value found for tritiated phases.
This fact will be relevant for the melting temperature of the classical
system.

The temperature dependence of the relative free energy of water and
ice at ambient pressure is presented in Fig. \ref{fig:10} for the
classical limit. The computational conditions for the non-equilibrium
RS simulations were summarized in the last two rows of Table \ref{tab:4}.
The estimated classical melting point is 259.3$\pm0.5$ K. We find
again an excellent agreement to the result reported by Habershon \emph{et
al.\citep{habershon09} }of 259$\pm1$K. Note that both classical
melting temperatures were determined by different methods, i.e., by
direct coexistence simulations\citep{habershon09} and by free energy
calculations. An advantage of the free energy method is that other
important physical quantities are readily available, which is not
the case by a direct coexistence method. 

In Table \ref{tab:5} we summarize the simulation results of several
melting properties for the studied isotopic compositions of water
and also in the classical limit. The melting entropy, $\Delta S_{m}$,
was estimated from the numerical temperature derivative of the Gibbs
free energy curves given in Figs. \ref{fig:6}-\ref{fig:8} and
\ref{fig:10}
at coexistence conditions, while the melting enthalpy was calculated
by two independent ways: \emph{(a)} as $T_{m}\Delta S_{m}$; \emph{(b)}
by direct calculation of the solid and liquid enthalpies by $NPT$
simulations at coexistence conditions. The agreement between both
methods provides evidence about the internal consistency of our
results.
The isotope effect in the melting entropy and enthalpy is low,
practically
within the statistical uncertainty of our results. The employed water
model underestimates both the experimental values of the melting
entropy
and enthalpy. At coexistence conditions the negative sign in the change
of the KE upon melting indicates that the KE of ice is larger than
that of the liquid phase.

\begin{widetext}

\begin{table}[t]
\caption{Melting temperature, entropy and enthalpy for normal, heavy,
and tritiated
water as well as classical limit results at ambient pressure. The
melting enthalpy was estimated by two independent methods. The change
in the kinetic and potential energy upon melting (liquid minus solid
values) and the molar volume ($V_{m}$) of the solid and liquid phases
are also given. The standard error in the final digits is given in
parenthesis. }
\label{tab:5}
\begin{tabular}{cccccc}
\hline
 & classical & T$_{2}$O & D$_{2}$O & H$_{2}$O & $\;$exp.
H$_{2}$O\tabularnewline
\hline
$T_{m}$(K) & 259.3(5) & 259.2(5) & 257.5(5) & 251 &
273.15\tabularnewline
$\Delta S_{m}$ (cal mol$^{-1}$K$^{-1}$) & 4.52(5) & 4.45(5) & 4.52(5) &
4.40(5) & 5.3\tabularnewline
$\Delta H_{m}$%
\footnote{from $T_{m}\Delta S_{m}$%
} (kcal mol$^{-1}$) & 1.17(2) & 1.15(2) & 1.16(2) & 1.10(2) &
1.44\tabularnewline
$\Delta H_{m}$%
\footnote{from independent solid and liquid $NPT$ simulations at
coexistence conditions%
} (kcal mol$^{-1}$) & 1.18(3) & 1.10(3) & 1.13(3) & 1.07(3) &
1.44\tabularnewline
$\triangle K_{m}$ ( kcal mol$^{-1}$ ) & 0 & -0.024(5) & -0.038(5) &
-0.078(5) & \tabularnewline
$\triangle U_{pot}$(kcal mol$^{-1}$) & 1.18(3) & 1.12(3) & 1.17(3) &
1.14(3) & \tabularnewline
$V_{m,s}$(cm$^{3}$mol$^{-1}$) & 19.40(1) & 19.47(1) & 19.49(1) &
19.56(1) & 19.66\tabularnewline
$V_{m,l}$(cm$^{3}$mol$^{-1}$) & 17.99(6) & 18.15(6) & 18.03(6) &
17.96(6) & 18.02\tabularnewline
$\triangle V_{m}$(cm$^{3}$mol$^{-1}$) & -1.41(6) & -1.32(6) & -1.46(6)
& -1.60(6) & -1.64\tabularnewline
\end{tabular}
\end{table}

\end{widetext}

At room temperature the KE of tritium in tritiated water is a factor
about 2.4 times larger than in the classical limit (see Table \ref{tab:1}),
i.e., although tritium is three times heavier than hydrogen, quantum
effects related to its nuclear mass are still significant below room
temperature. Therefore an unexpected result of the employed model
potential is that the classical melting point is nearly identical
to that one found for tritiated water, and the reason for this behavior
will be investigated in the next subsection.

\subsection{Kinetic energy and molecular mass\label{subsec_kinetic} }

\begin{figure}
\vspace{-1.3cm}
\hspace*{-0.1cm}
\includegraphics[width= 9cm]{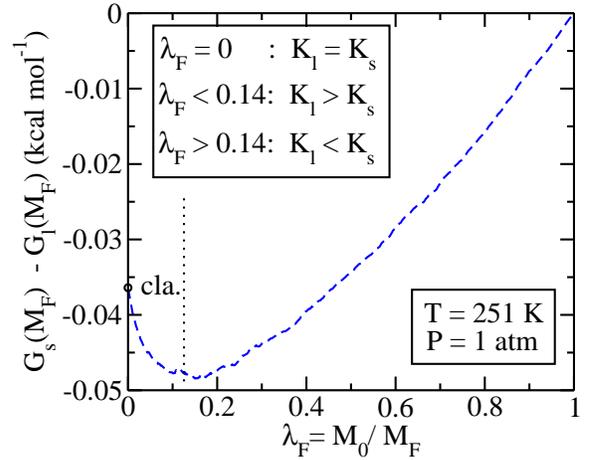}
\vspace{-1.3cm}
\caption{Gibbs free energy difference between ice and water as a
function of
the molecular mass at the reference state point ($T_{R},P_{R}$).
Depending on the molecular mass the KE of the liquid ($K_{l}$) may
be larger than that of the solid ($K_{s}$).}
\label{fig:11}
\end{figure}

We have already seen that isotope shifts in the melting point are
caused by the fact that the free energy of the solid and liquid phases
changes by different amounts as the isotopic mass changes. This is
a quantum effect related to the atomic mass, as in the classical limit
the mass dependence of the free energy is identical for both phases.
To understand the shift found in the melting temperature in the classical
limit, it is interesting to study the difference between the Gibbs
free energy of the solid and liquid phases as a function of the scaled
molecular mass, $M_{F}$, as determined by the parameter $\lambda_{F}=M_{0}/M_{F}$.
This difference is given by an expression similar to Eq. (\ref{eq:delta_G_cla})
\begin{eqnarray}
\hspace{-3cm}
G_{R,s}(M_{F})-G_{R,l}(M_{F}) =  \hspace{3.5cm}  \nonumber    \\
= \int_{1}^{\lambda_{F}}\left(\frac{\left\langle K_{s}(M)\right\rangle _{NPT}}{\lambda}-\frac{\left\langle K_{l}(M)\right\rangle _{NPT}}{\lambda}\right)d\lambda\;.
\label{eq:delta_G_M}
\end{eqnarray}
The instantaneous values of the virial KE estimators of the integrand
are available from the non-equilibrium AS simulations used to calculate
Fig. \ref{fig:9}. These values have been used to plot the solid-liquid
Gibbs free energy difference as a function of the inverse molecular
mass, $M_{F}^{-1}$, in Fig. \ref{fig:11}. This curve is not a monotonous
function but it displays a minimum for a molecular mass defined by
the parameter $\lambda_{F}\sim0.14$. This minimum makes that the
free energy difference in the classical limit (corresponding to $\lambda_{F}=0$,
open circle in Fig. \ref{fig:11}) is similar to that obtained for
tritiated phases (see Table \ref{tab:3}), and therefore the classical
melting point is found at nearly the same temperature as for T$_{2}$O.
If the curve in Fig. \ref{fig:11} were a monotonous decreasing function
up to $\lambda_{F}=0$ (i.e., without the presence of a minimum),
then the classical absolute value of the free energy difference between
the two phases should be larger than for T$_{2}$O, and therefore
the classical melting temperature should also increase with respect
to the tritiated phase.

\begin{figure}
\vspace{-1.8cm}
\hspace*{-0.1cm}
\includegraphics[width= 9cm]{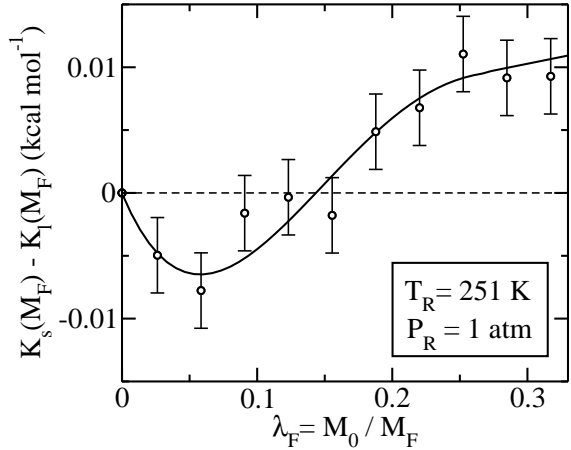}
\vspace{-1.3cm}
\caption{Kinetic energy difference between ice and water as a function
of the
molecular mass at the reference state point ($T_{R,}P_{R}$). The
line is a guide to the eye.}
\label{fig:12}
\end{figure}

\begin{figure}
\vspace{-1.8cm}
\hspace*{-0.1cm}
\includegraphics[width= 9cm]{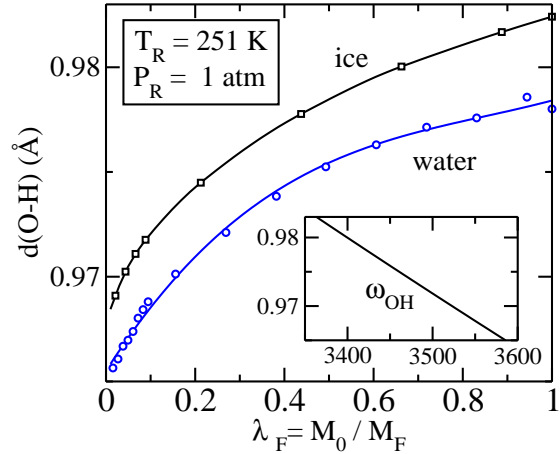}
\vspace{-1.3cm}
\caption{Intramolecular OH distance of ice and water as a function of
the molecular
mass at the reference state point ($T_{R,}P_{R}$). Lines are guides
to the eye. The inset shows the quasi-harmonic stretch frequency
$\omega_{OH}$
(in cm$^{-1}$) as a function of the OH distance for the q-TIP4P/F
potential. $\omega_{OH}$ was derived from the second derivative of
the potential energy with respect the OH bond distance and by
considering
the actual O and H masses.}
\label{fig:13}
\end{figure}

The existence of a minimum in the function given in Fig. \ref{fig:11}
is related to a change in the sign of the integrand of Eq. (\ref{eq:delta_G_M})
in the region around the minimum. This fact is clearly seen by plotting
in Fig. \ref{fig:12} the difference between the KE of solid and liquid
phases as a function of the inverse molecular mass, $M_{F}^{-1}$,
at the reference state point. We find that when the molecular mass
is large ($\lambda_{F}<0.14,$ or $M_{F}>7M_{0}$ ) the KE of water
is larger than that of ice, while the opposite behavior is found for
lower molecular masses ($\lambda_{F}>0.14$ ). The physical origin
of this apparently complicate behavior of the KE is related to the
presence of the two types of bonds in the water phases: intramolecular
OH pairs and intermolecular H-bonds. The H-bond network is stronger
in ice than in water, as a result of the higher molecular disorder
in the liquid, which implies that vibrational frequencies of H-bonds
are larger in ice than in water. On the contrary, the OH stretch frequency
in the liquid is higher than in the solid. In Fig. \ref{fig:13} we
display the OH distance for water and ice as a function of the molecular
mass at the reference state point. For all molecular masses, the OH
distance is lower for water than for ice, a fact that implies also
a higher vibrational OH stretch frequency in water. This behavior
is illustrated by the inset in Fig. \ref{fig:13} that displays the
increase of the quasi-harmonic stretch frequency $\omega_{OH}$ for
the employed q-TIP4P/F potential as the OH distance decreases. 

In the classical limit, corresponding to $\lambda_{F}=0$ (infinite
molecular mass), all modes have the same KE (equipartition principle)
and the KE difference between ice and water vanishes. In the case
of large, but finite, molecular mass ($0<\lambda_{F}<0.14$) , the
KE difference is determined by the modes with highest vibrational
frequencies, i.e., the OH stretches. The reason is that the leading
quantum correction for vibrational modes, as the molecular mass decreases,
corresponds to those modes that satisfy that their energy quantum
is larger than the thermal energy, $\hslash\omega>k_{B}T_{R}$, and
this condition will be first met for the modes with highest frequency
(OH stretches). Thus, for large molecular masses with $\lambda_{F}<0.14$,
the KE of water results larger than that of ice, as the stretch frequencies
$\omega_{OH}$ are larger in the liquid. For smaller molecular masses
($\lambda_{F}>0.14$), the quantum behavior of the vibrational modes
associated to H-bonds becomes important. For H-bonds related modes,
the KE is larger in ice than in water, as their vibrational frequencies
are higher in ice. The effect of the H-bonds dominates over the OH
stretches for $\lambda_{F}>0.14$ in the sense that the KE of ice
results larger than that of water.

The shorter OH distance in liquid water is a consequence of the coupling
between OH stretches and H-bonds. Since the H-bond network in water
is weaker than in ice, then the intramolecular OH bonds become shorter.
A recent Compton scattering study of water versus ice Ih arrives to
the same conclusion: an elongation of the H-bond in water leads to
a systematic shortening of the intramolecular OH bond, and the OH
bond in water is about 0.01 \AA\ shorter than in ice.\citep{nygard06}

\section{Conclusions\label{sec:conclusions}}

In the present work we have used free energy techniques to study the
isotope shift in the melting temperature of ice Ih. The employed AS
and RS methods are based on algorithms where the Hamiltonian or a
state variable are adiabatically changed along a simulation run. The
reversible work associated to this change is equal to the free energy
difference between the initial and final state, as long as the change
is performed slowly (adiabatically). These free energy algorithms
are easily implemented in any code prepared for equilibrium simulations.

Our Pl simulations of water have been done with the q-TIP4P/F model,
a point charge potential that includes an anharmonic treatment of
molecular flexibility. Several equilibrium properties (RDFs, liquid
density as a function of temperature) have been compared to results
previously published for this potential. The agreement found between
independent simulations provides a check for the employed computational
method and simulation conditions. 

We have found that the experimental isotope effect in the TMD of water
is not correctly described by the employed potential. In fact, the
TMD calculated in the classical limit is very close to the quantum
value. Quantum effects associated to the atomic mass activate a competing
mechanism that produces a weakening of the  H-bond through quantum
zero point fluctuations, but also a strengthening of the same H-bonds
by the increase in the molecular dipole moment of water molecules,
that is found in the quantum simulations. Although this competing
mechanism, previously observed by Habershon \emph{et al.}\citep{habershon09}
to explain the trends in the classical and quantum diffusion coefficients,
seems to be a real physical effect, its influence in a given physical
property depends on the details of the anharmonic flexible water potential,
so that a small imbalance between both competing factors might lead
to an unphysical result. This seems to be the case for the isotope
effect in the TMD of water.

The isotope effect in the melting temperature at ambient pressure
of ice has been determined by the calculation of the Gibbs free energy
as a function of the isotope mass and temperature. We find that the
isotope effect predicted by the q-TIP4P/F potential (6.5$\pm$0.5
K and 8.2$\pm$0.5 K for heavy and tritiated water, respectively)
are larger than the experimental values (3.8 K and 4.5 K, respectively).
An unexpected result is that the classical melting point at 1 atm
is nearly identical to the one obtained for tritiated water. We have
shown that this behavior is related to the fact that the OH stretches
in water display a higher frequency than in the solid phase. It is
clear that the employed q-TIP4P/F potential is not able to quantitatively
predict the isotope effect in the melting temperature of ice, but
it has helped to identify that the coupling between molecular flexibility
and the H-bond network may have significant implications in the phase
behavior of water. 

\acknowledgments 

This work was supported by Ministerio de Ciencia e Innovaci\'on (Spain)
through Grant No. FIS2009-12721-C04-04 and by Comunidad Aut\'onoma de
Madrid through project MODELICO-CM/S2009ESP-1691. The authors benefited
from discussions with L. M. Ses\'e, C. Vega, E.G. Noya, and E. R.
Hern\'andez.

\vspace*{1.0cm}

\appendix

\section{Extrapolation of $G_{R,cla}$\label{appendix: harmonic}}

The largest deviation between a quantum and classical treatment of
a water phase (either solid or liquid) is expected to be due to those
vibrational modes of highest frequency (i.e., intramolecular OH stretches
and H-bonds). A harmonic treatment of these modes gives useful information
on the analytical behavior of the definite integral in Eq. (\ref{eq:delta_G_M})
when the upper integration limit vanishes ($\lambda_{F}\rightarrow0$).
Note that the limit $\lambda_{F}=0$ is not accessible numerically
as it would correspond to an infinite mass. Let us assume a simple
one-dimensional harmonic oscillator defined by a wavenumber $\omega_{0}$
and mass $M_{0}$. The oscillator mass depends on the integration
parameter $\lambda$ as $M=M_{0}/\lambda$, while the oscillator wavenumber
varies as $\omega=\omega_{0}\lambda^{1/2}$. The thermal expectation
value of the KE of the quantum oscillator for a given $\lambda$ is
given by

\begin{equation}
\left\langle K\right\rangle =K_{0}\lambda^{1/2}\coth\left(\frac{K_{0}}{K_{cla}}\lambda^{1/2}\right)\;,\label{eq:K_cua}\end{equation}
where $K_{cla}$ is the classical KE

\begin{equation}
K_{cla}=\frac{1}{2}k_{B}T\;,\end{equation}
and $K_{0}$ is the zero-point KE 

\begin{equation}
K_{0}=\frac{\hslash\omega_{0}}{4}\;.\end{equation}
By a Taylor expansion of the r.h.s. of Eq.(\ref{eq:K_cua}), the integrand
of Eq. (\ref{eq:delta_G_cla}) for a harmonic mode can be written
as

\begin{equation}
\frac{\left\langle K\right\rangle -K_{cla}}{\lambda}=\frac{K_{0}^{2}}{K_{cla}}-\frac{K_{0}^{4}}{K_{cla}^{3}}\lambda+O(\lambda^{2}\text{)}\;.\label{eq:taylor}\end{equation}
The definite integral given in Eq. (\ref{eq:delta_G_cla}) must be
obtained by numerical extrapolation of $I(\lambda_{F})$ to the $\lambda_{F}=0$
limit {[}see Eq. (\ref{eq:I_lim}){]}. The harmonic result for the
integrand in Eq. (\ref{eq:taylor}) suggests that a polynomial fit
for $I(\lambda_{F})$ is a convenient numerical way to obtain the
desired extrapolation. As the employed q-TIP4F/P potential is anharmonic,
we do not expect to assign any particular meaning to the fitted coefficients.

\end{document}